\documentclass[english,aps,manuscript,prl, twocolumn,reprint]{revtex4-1}
\usepackage[latin9]{inputenc}
\usepackage{amsmath}
\usepackage{graphicx}
\usepackage{xargs}[2008/03/08]
\usepackage{color}
\makeatletter

\@ifundefined{textcolor}{}
{%
 \definecolor{BLACK}{gray}{0}
 \definecolor{WHITE}{gray}{1}
 \definecolor{RED}{rgb}{1,0,0}
 \definecolor{GREEN}{rgb}{0,1,0}
 \definecolor{BLUE}{rgb}{0,0,1}
 \definecolor{CYAN}{cmyk}{1,0,0,0}
 \definecolor{MAGENTA}{cmyk}{0,1,0,0}
 \definecolor{YELLOW}{cmyk}{0,0,1,0}
}

\usepackage{polski}

\makeatother

\usepackage{babel}
\begin{document}

\newcommandx\ket[1][usedefault, addprefix=\global, 1=]{|#1\rangle}

\newcommandx\bra[1][usedefault, addprefix=\global, 1=]{\langle#1|}

\newcommandx\avg[1][usedefault, addprefix=\global, 1=]{\langle#1\rangle}

\newcommandx\var[1][usedefault, addprefix=\global, 1=]{\langle(\Delta#1)^{2}\rangle}

\global\long\def\aa{\hat{a}}

\global\long\def\ad{\hat{a}^{\dagger}}

\title{Hologram of a Single Photon}

\author{Rados\l{}aw Chrapkiewicz}
\author{Micha\l{} Jachura}
\email{michal.jachura@fuw.edu.pl}
\author{Konrad Banaszek}
\author{Wojciech Wasilewski}

\affiliation{Faculty of Physics, University of Warsaw, Pasteura 5, 02-093 Warsaw, Poland}

\date{\today}

\maketitle

\textbf{
The spatial structure of single photons \cite{Molina-Terriza2007b,Lundeen2011,Fickler2012} is becoming an extensively explored resource 
used for 
facilitating the free-space quantum key distribution \cite{Walborn2006a,Wang2012a,Vallone2014,Krenn2015} and
quantum computation \cite{Abouraddy2012b} as well as for 
benchmarking the limits of quantum entanglement generation \cite{Fickler2012}
with orbital angular momentum modes \cite{Molina-Terriza2007b,Nagali2009a} or
 reduction of the photon free-space propagation speed \cite{Giovannini2015}. 
Albeit nowadays an accurate tailoring of photon's spatial structure is routinely performed using methods employed for shaping classical optical beams \cite{Dholakia2011a,Fickler2012,Giovannini2015},
the reciprocal problem of retrieving the spatial phase-amplitude structure of an unknown single photon cannot be solved using complimentary classical holography techniques \cite{GABOR1948,Collier1971}  exhibiting excellent interferometric precision. Here we introduce a method to record a hologram of a single photon (HSP) probed by another reference photon, based on essentially different concept of quantum interference between two-photon probability amplitudes. Similarly to classical holograms, HSP encodes full information about photon's ``shape'', i.e. its quantum wavefunction whose local amplitude and phase are retrieved in the demonstrated experiment. }

The complete characterization of a quantum wavefunction of an unknown photon presents a challenging task whose difficulty lies particularly in the retrieval of its local phase variations. This is caused by the fundamental property of single photons i. e. their entirely indeterminate global phase following
from the perfect rotational symmetry of their Wigner functions in the phase space \cite{lvovsky2001quantum}, which precludes the application of interferometric techniques such as optical holography utilizing fixed phase relation between investigated and reference light. Therefore the characterization of photon's spatial structure has never benefited from the precision and a simplicity provided by the holography methods \cite{GABOR1948,Collier1971} but as yet has been tackled only using indirect tomographic techniques \cite{Smith2005a} or weak values measurements \cite{Lundeen2011}.

\begin{figure}[!bh]
\includegraphics[width=0.95\columnwidth]{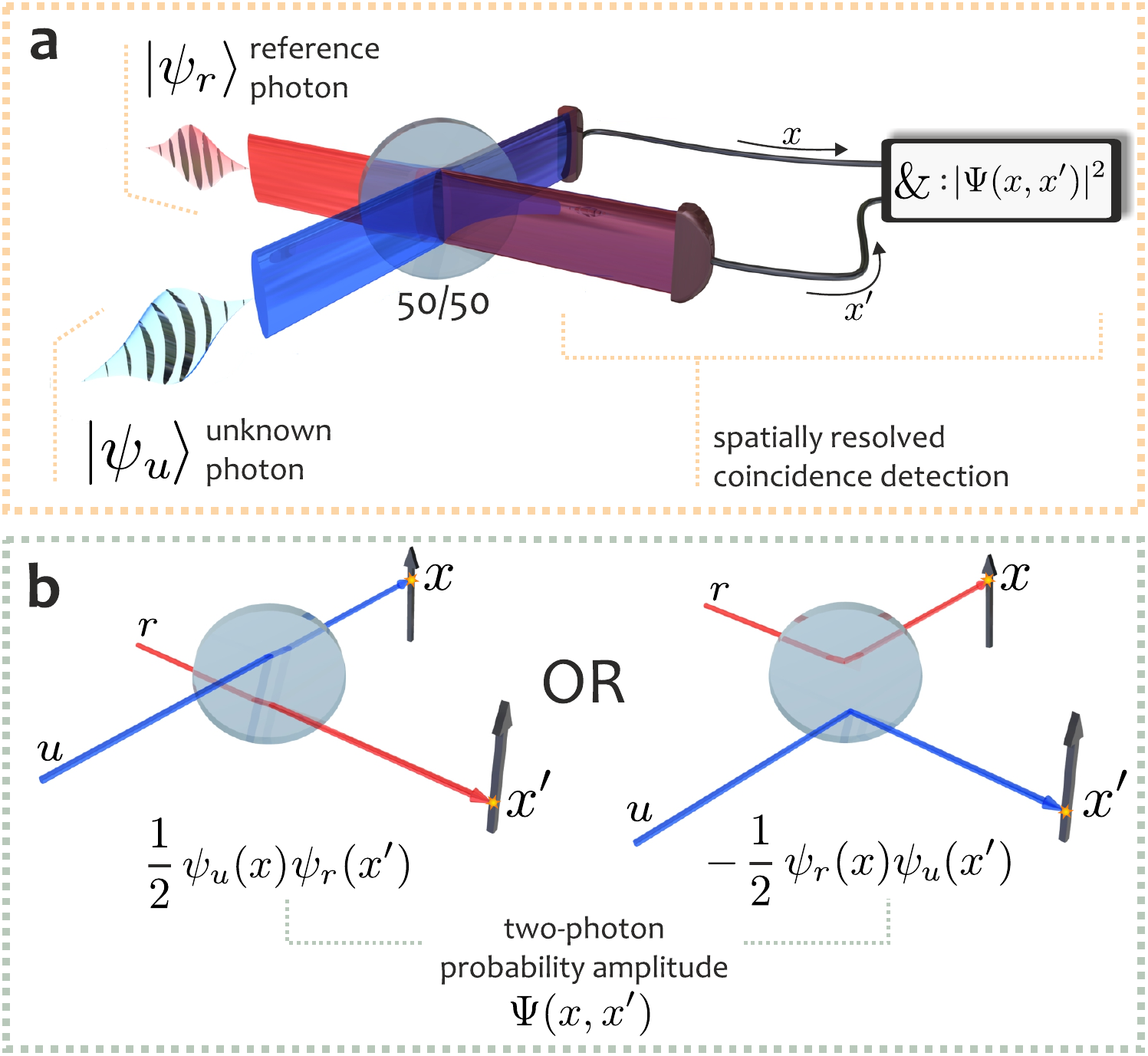}

\protect\caption{\textbf{Quantum interference of two spatially structured photons.
} \textbf{a}, In analogy to
classical holography we repeatedly overlap an unknown photon
$\protect\ket[\psi_{u}]$ with a reference (known) photon $\protect\ket[\psi_{r}]$ with the constant local phase profile on a 50/50 beam splitter and we spatially localize coincidence events in $x$ and $x'$, measuring their joint probability distribution $|\Psi(x,x')|^{2}$ which is sensitive to any differences between the quantum wavefunctions of
the photons $\psi_{u}(x)$ and $\psi_{r}(x)$ including the local
variations of their phases. \textbf{b}, The spatially localized coincidence events $(x,x')$ originate from the non-destructive interference of probability amplitudes of
two classically exclusive, but quantum-mechanically coexisting
scenarios: (left) the unknown photon in $x$ and the
reference photon in $x'$ have passed through the beam splitter,
(right) both photons localized conversely in $x'$ and $x$ have been
reflected from the beam splitter. }
\end{figure}
\begin{figure*}[t]
\includegraphics[width=0.9\textwidth]{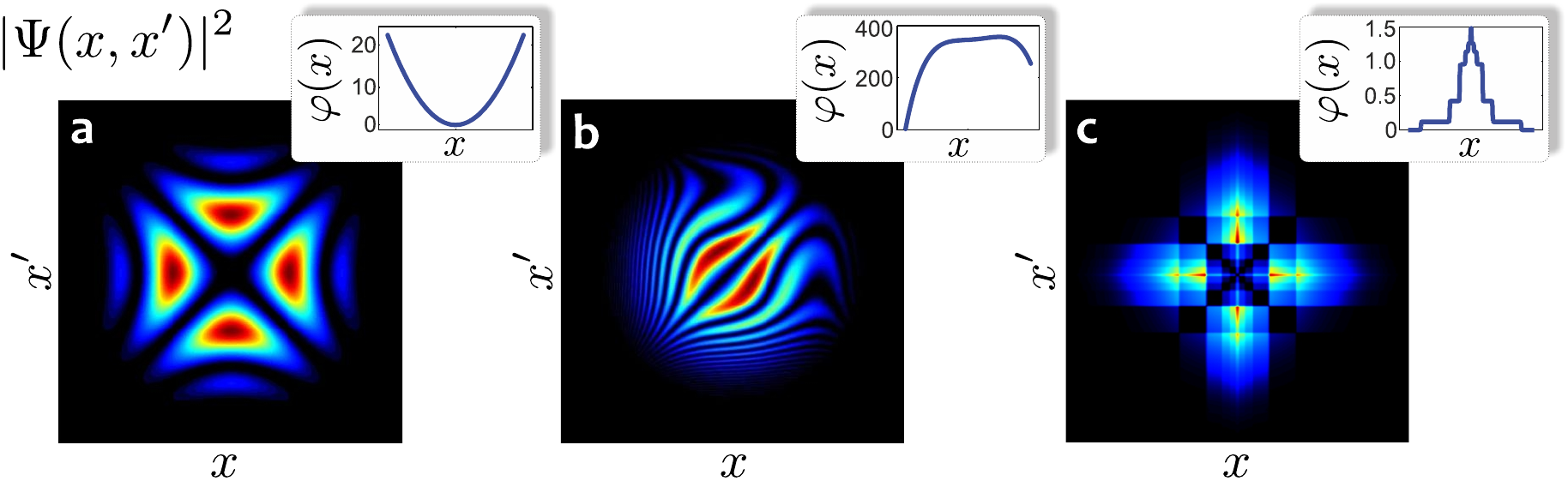}

\protect\caption{\textbf{Encoding of the local phase of quantum wavefunction in the hologram of a single photon (HSP).} The HSP emerging from a joint probability distribution
of the coincidence events $|\Psi(x,x')|^{2}$ encodes (see equation (\ref{eq:pxx}))
the local phase profile of the unknown photon $\varphi(x)$. \textbf{a-c}, To illuminate
this feature originating from the local phase sensitivity
of the quantum interference, we depict the expected HSP structure
(false colors denotes computed probability gradations) for photons
in two gaussian modes with identical amplitudes $|\psi_{u}(x)|=|\psi_{r}(x)|$,
differing by the local phase profile of the unknown photon presented in the upper-right
corner of each plot. \textbf{a}, For the experimental demonstration the purely
quadratic local phase profile has been chosen. \textbf{b}, HSP for the generic fourth-order
polynomial local phase profile. \textbf{c}, HSP for the non-polynomial local phase profile resembling
a fragment of the Warsaw skyline.}
\end{figure*}
In this paper, we experimentally show that the hologram of a single photon (HSP)
encoding full information about its spatial structure given by the
quantum wavefunction $\psi(x)=\langle x|\psi\rangle$ \cite{Lundeen2011}
can be recorded if the first-order interference of optical
fields is replaced by the non-classical interference of spatially
varying two-photon probability amplitudes. The idea of HSP, sketched
in Fig.~1a, relies on overlapping the unknown photon $\ket[\psi_{u}]$ of an arbitrary local phase profile $\varphi(x)=\mathrm{arg}(\psi_{u}(x))$ with a reference photon $\ket[\psi_{r}]$ having the constant local phase profile on a beam splitter, both photons occupying similar spectral (temporal) modes. Afterwards we measure positions of photons which coincidentally left two distinct output ports of the beam splitter parametrized by $x$ and $x'$ coordinates. Any
feature distinguishing photons, such as local difference between their quantum
wavefunctions $\psi_{u}(x)$ and $\psi_{r}(x)$ prevents them from
ideal two-photon coalescence known as Hong-Ou-Mandel effect \cite{Hong1987},
thus the observation of spatially localized coincidences $(x,x')$
serves as a sensitive probe of the spatial structure of the unknown
photon. As we visualize in Fig. 1b, such a coincidence event can
originate either from transmission or reflection of both photons at
the beam splitter. These two fundamentally indistinguishable events
account simultaneously to a two-photon probability amplitude $\Psi(x,x')$ describing
one photon localized at position $x$ and the other at $x'$, which
can be expressed in Feynman's path integral formalism as:
\begin{figure}[b]
\includegraphics[width=1\columnwidth]{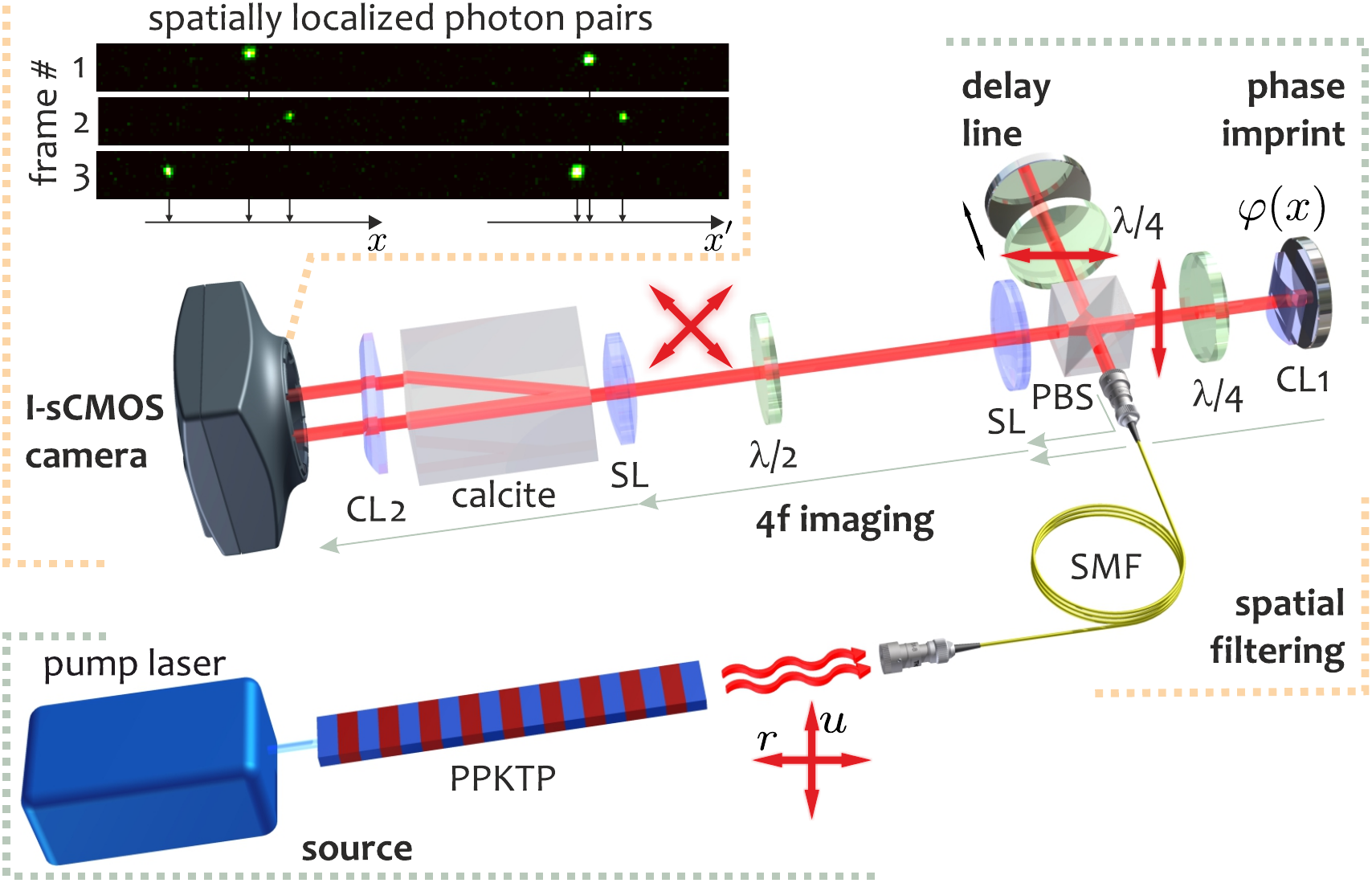}
\centering
\protect\caption{\textbf{Experimental setup for measuring HSP. } Orthogonally polarized unknown and the reference photon, generated in spontaneous parametric down conversion process, are prepared in the same spectral mode. The photons are
transmitted through the single mode fiber (SMF), separated by the polarization
beam splitter (PBS) and then, at the output beam waist, the local phase profile $\varphi(x)$ is imprinted on the unknown photon during its double pass propagation through a phase mask (a cylindrical lens (CL1) for the quadratic phase as in Fig. 2a). We localized photons outgoing from two distinct
ports of a beam-splitter, here implemented collinearly as a half wave plate ($\lambda/2$)
and calcite crystal, by  means of the state-of-the-art intensified
sCMOS camera  \cite{Chrapkiewicz2014a,Jachura2015c}. Both the beam waist surface
of the reference photon and the phase mask surface were mapped onto
the camera with a phase-preserving $4f$ system consisting of two spherical lenses (SL). }
\end{figure}
\begin{figure*}[t]
\includegraphics[width=0.7\paperwidth]{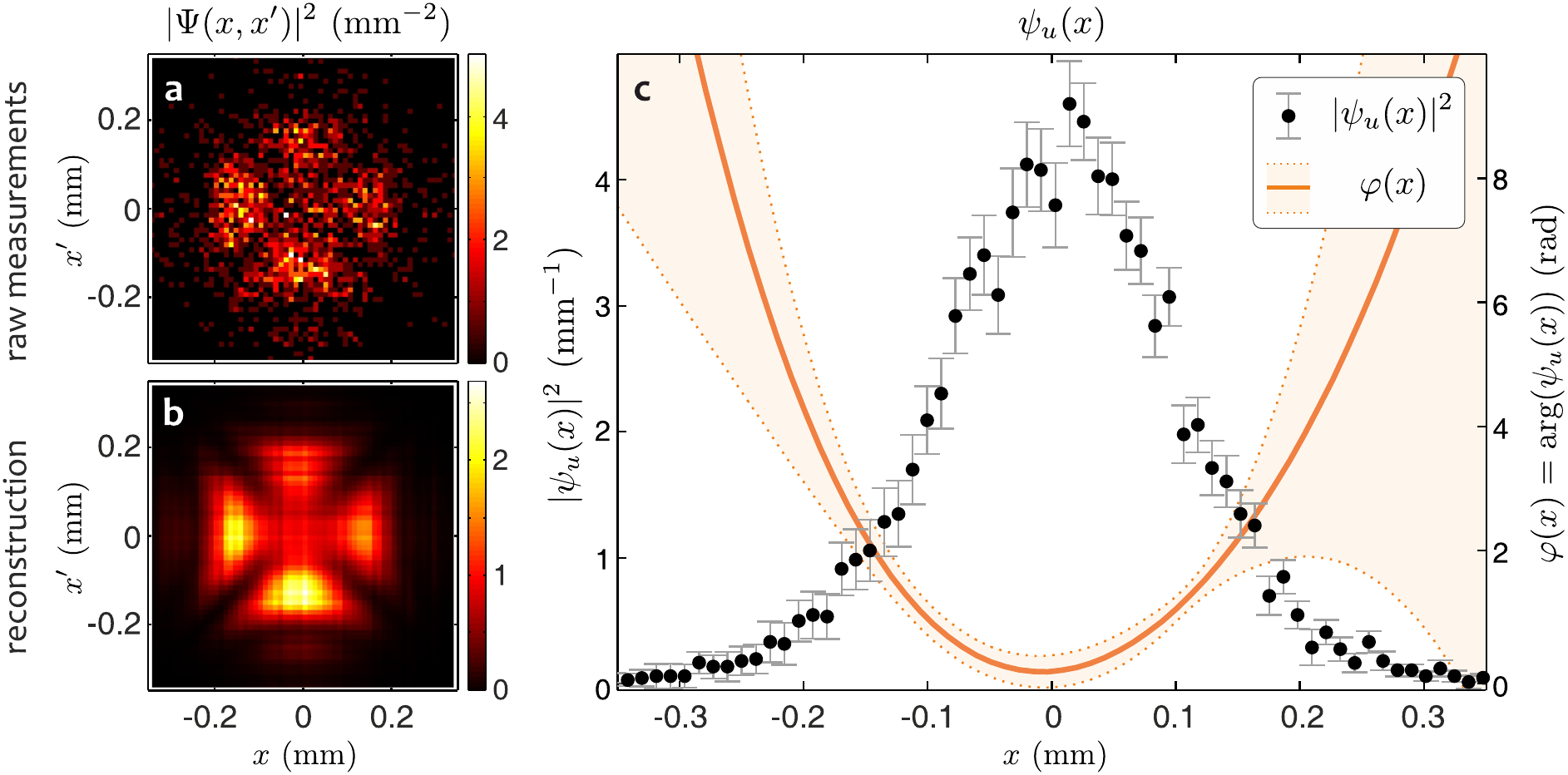}

\protect\caption{\textbf{Measured and reconstructed HSP along with the full retrieval
of encoded quantum wavefunction.} \textbf{a}, Directly measured joint probability distribution $|\Psi(x,x')|^{2}$ form an empirical HSP. \textbf{b}, Utilizing independently measured, nearly identical amplitudes of the quantum wavefunctions $|\psi_{u}(x)|$, $|\psi_{r}(x)|$ we numerically
reconstructed HSP best matching the raw experimental data which closely resembles the theoretically predicted pattern presented in Fig. 2a. \textbf{c}, The measurements followed by the numerical reconstruction yields the complex
quantum wavefunction of the unknown photon $\psi_{u}(x)$, in particular its
phase $\varphi(x)=\arg(\psi_{u}(x))$. Uncertainty ranges stand for one standard deviation  (see Methods for details). }
\end{figure*}

\begin{equation}
\Psi(x,x')=\frac{1}{2}(\langle x|\psi_{u}\rangle\langle x'|\psi_{r}\rangle-\langle x|\psi_{r}\rangle\langle x'|\psi_{u}\rangle).\label{eq:probamp}
\end{equation}

Thanks to the recent advances in spatially resolved detection techniques such as fiber-coupled avalanche photodiodes \cite{Peeters2009,Rozema2014} or single photon sensitive intensified cameras \cite{Chrapkiewicz2014a,Jachura2015c} we were able to measure the joint probability distribution $|\Psi(x,x')|^{2}$ with the resolution high enough to reveal its spatial variations originating from the non-destructive interference of the unknown and the reference
photon quantum paths. Remarkably this joint probability distribution provides information about the local phase profile of the unknown photon $\varphi(x)$:

\begin{multline}
|\Psi(x,x')|^{2}=\frac{1}{4}(|\psi_{u}(x)|^{2}|\psi_{r}(x')|^{2}+|\psi_{r}(x)|^{2}|\psi_{u}(x')|^{2})-\\
\frac{\mathcal{V}}{2}|\psi_{u}(x)||\psi_{u}(x')||\psi_{r}(x)||\psi_{r}(x')|\cos(\varphi(x)-\varphi(x')).\label{eq:pxx}
\end{multline}
The HSP given by $|\Psi(x,x')|^{2}$ is entirely insensitive to any constant offset of the local phase profile of the unknown photon in contrast to optical holograms which, while being recorded, are extremely sensitive to a phase shift between reference and unknown fields. The visibility of the HSP fringes $\mathcal{V}$ is  defined by a spectral (temporal) mode
overlap which can be high and stable for photons generated by different
sources such as two independent spontaneous parametric down-conversion
(SPDC) sources \cite{Kaltenbaek2006}, quantum dots \cite{Patel2010}
or even dissimilar sources \cite{Bennett2009}. In Fig. 2 we visualize the HSP structures for exemplary local phase profiles. We can resort to one of the numerous methods of phase retrieval \cite{Servin:97} to infer $\varphi(x)$ from equation (\ref{eq:pxx}), as the detection probability distributions $|\psi_{u}(x)|^{2}$, $|\psi_{r}(x)|^{2}$ are directly measurable quantities.

We selected for experimental demonstration of HSP the situation
depicted in Fig. 2a where the unknown photon has the quadratic local
phase $\varphi(x)=kx^{2}/2R$ resulting in a cross-shaped HSP. Here
$R$ stands for a radius of  curvature, and $k=2\pi/800$~nm for the
wave number. Both the unknown and the reference photons were generated
via type-II spontaneous parametric down conversion process realized in a periodically poled KTP \mbox{(PPKTP)} nonlinear crystal pumped with 400 nm light from a continuous wave diode laser. We ensured the high indistinguishability of their spectral (temporal) modes, confirmed in an independent HOM dip measurement yielding the visibility of 91\%. 

As presented in Fig. 3, the photons were spatially filtered by a single mode fiber and then separated by polarization beam splitter and directed separately to two arms of a delay line and phase imprinting system. The lengths of the arms were adjusted
to overlap photons temporally and to set the constant-phase waists of
the mode coming out from a fiber collimator on the mirrors surfaces.
We inserted a cylindrical lens ($f_c=75\,\mathrm{mm}$) in the proximity
of one of the mirrors, thus imprinting the quadratic local phase profile in
a horizontal direction on the unknown photon during its back and forth
propagation. Since the reference and unknown photon propagating through different arms of the delay line were orthogonally polarized, no interference occured at the delay line output.    

The key part of the setup was an
intensified complementary metal-oxide semiconductor (I-sCMOS) camera system of parameters suitable to detect spatially
resolved photon pairs (see refs \cite{Chrapkiewicz2014a,Jachura2015c} and Methods for camera operation details).
We imaged the delay-line mirror surfaces on the camera using a $4f$ system
preserving both amplitude and the phase of impinging photons spatial
wavefunctions. A cylindrical lens (CL2) placed in front of the camera reduced
the mode size in the vertical direction perpendicular to the plane
of the setup and consequently frame reading time.

The beam splitter transformation was implemented in the collinear
configuration as a half wave plate followed by calcite polarization displacer such that its two output ports corresponded to the two distinct regions of the camera. In the experiment we retained for analysis frames containing two detected photons, registering their positions in a horizontal dimension parametrized by $x,x'$ coordinates in respective regions of sCMOS sensor. High spatial resolution allowed us to record the subtle variations of the detected photons positions and thus directly measure the empirical coincidence probability distribution $|\Psi(x,x')|^{2}$. 

The measured HSP consisting of approximately $2.2\times10^{3}$ detected
photon pairs is presented in Fig. 4a, which closely resembles the
theoretically predicted cross-like shape shown in Fig. 2a. Following
equation (\ref{eq:pxx}) we decoded the phase $\varphi(x)$ 
using one of a  numerical methods (see ref  \cite{Servin:97} and Methods for details), which finds
the local phase profile that yields the coincidence probability distribution
closest to the measured data as displayed in Fig. 4b. The procedure
was fed with virtually identical wavefunctions amplitudes of the unknown photon $|\psi_{u}(x)|$, presented in Fig. 4c, and the reference photon $|\psi_{r}(x)|$ measured independently using the coincidence imaging scheme (see ref \cite{Jachura2015c} and Methods for details). 

We show the complex quantum wavefunction of an unknown photon i.e. its measured
amplitude and the phase extracted from its HSP along with the uncertainty
ranges in Fig. 4c. We found the radius of curvature of the reconstructed
local phase profile of the photon $R=34\pm1.5$ mm to be in a good
agreement with the value expected from a double pass through the phase
imprinting lens which has been confirmed in an independent measurement
by interfering a classical beams in this setup. The uncertainty
of the reconstructed phase is below $2\pi/25$ in a central region
and it diverges only on the edges of the wavefunction due to the scarcity
of registered counts outside the central region.

The HSP method naturally transfers the hologram recording techniques into the field of quantum optics presenting a compelling and promising way of the quantum wavefunction retrieval. The technique can be readily adapted to the more general configurations where a reference photon has unknown structure by spatially shearing \cite{Walmsley2009} the photons in the second measurement run (see Supplementary Information for details). HSP technique can be also extended to the two-dimensional case requiring the efficient detection of a four dimensional coincidence probability distribution (see Supplementary Information for details). Parallel development of low-jitter, time-resolving detectors
would allow to readily implement HSP in the mathematically equivalent
spectral (temporal) domain where local phase sensitivity of the non-classical interference has been observed \cite{Specht2009}, and several wavefunction reconstruction techniques have been presented \cite{Wasilewski2007,Beduini2014,Polycarpou2012}. 
Finally let us emphasize that since our scheme relies solely on multiparticle bosonic interference, it can be generalized for all bosons. Prospective measurement of the hologram of a single atom and further retrieval of its wavefunction could utilize the scheme recently reported in the first experimental realization of two-boson interference \cite{Lopes2015} relying on a similar detection technique. 

\footnotesize{
\subsection*{Methods}

\textbf{Photon source.}
The photon pairs consisting of the unknown and the reference photon
were generated via type-II degenerate spontaneous parametric down-conversion
(SPDC) process realized in 5-mm long periodically poled KTP crystal
(poling period $9.2\,\mu\mathrm{m}$) pumped with 8 mW of 400 nm light
from a single mode, continuous wave diode laser. The temperature
of the crystal was stabilized to $24.1^{\circ}\mathrm{C}$ to
ensure maximal and stable overlap between the spectral modes of generated
photons. The photons were spectrally filtered by a narrowband 3-nm
full width at half-maximum interference filter, spatially filtered by a single
mode fiber,  and temporally overlapped after polarization beam splitter by means of an optical delay line 
where double pass through quarter wave plate ($\lambda/4$) rotated photons' polarization by $90^\circ$.

 We characterized the indistinguishability of photons used
in the experiment with the standard avalanche photodiode coincidence
system by measuring the Hong-Ou-Mandel dip yielding the visibility
of 91\%. 

\textbf{Single photon localization with I-sCMOS camera.}
To localize photons with high spatial resolution we used a camera system consisting of
scientific complementary metal-oxide semiconductor endowed with image intensifier camera (I-sCMOS),
assembled in our group. The image  intensifier begins with gallium arsenide
photocathode converting the impinging photons into electrons with
the quantum efficiency of $20\%$. Afterwards each electron enters
the multichannel plate where it triggers the growing charge avalanche
which hits the phosphor screen resulting in a bright green-light flash
of a decay time below 200 ns. A typical phosphor flash has a diameter
of $66$ $\mu$m and highly random brightness determined by the stochastic
avalanche process. The flashes are imaged on the sCMOS camera sensor
via a bright relay lens and real-time localized by a software algorithm, which retrieves central positions of the flashes from a raw image with subpixel accuracy.  We acquired a data from $1000\times20$ pixel region of interest selected on the sCMOS camera sensor corresponding to approximately $11.5\times10^{4}$ microchannels with a frame rate of 7 kHz. We set the time gate of the image intensifier to 30 ns ensuring that virtually no accidental coincidences or more than two photons per frame were detected. Moreover, the chosen time gate corresponded to a dark count rate of $4\times10^{-7}$ per microchannel which could be neglected in further analysis. See Supplementary Information for further details of camera construction and operation.

\textbf{Measurement details.}
We measured the amplitude of the wavefunctions $|\psi_{u,r}(x)|$
by setting the half wave plate $\lambda/2$ (HWP) to $\theta=0^{\circ}$ and $\theta=45^{\circ}$
to direct photons into different output ports of the calcite displacer.
The nearly identical squared amplitudes of both photons were recovered by directly
following our coincidence imaging scheme \cite{Jachura2015c}. Then we proceed
to the HSP measurement by setting HWP to $\theta=22.5^{\circ}$ interchangeably
measuring wavefunction amplitudes and HSP by rotating HWP after each
$5\times10^{6}$ frames out of $1.8\times10^8$ of total number of  collected frames.

\textbf{Phase-retrival algorithm.}
We found efficient to apply a numerical search of $\varphi(x)$ that
matches the measured HSP best, by solving the optimization problem
according to the general idea suggested by \cite{Servin:97}. We
performed the optimization procedure: 
\begin{equation*}
\min_{\varphi(x)}\big|\big||\Psi(x,x')|^{2}-|\Psi_{\textrm{rec}}^{(\varphi)}(x,x')|^{2}\big|\big|,\label{eq:optimal}
\end{equation*}
where $|\Psi(x,x')|^{2}$ stands for the measured empirical distribution, $|\Psi_{\textrm{rec}}^{(\varphi)}(x,x')|^{2}$ is a functional defined by equation~(2) constructed from the measured amplitudes $|\psi_{u,r}(x)|$ depending on the vectorized phase profile $\varphi(x)$ to be found, and $||\cdot||$ is the Frobenius norm of the matrix. Since the general global search is a computationally hard problem, we divided our optimization
into two simpler subsequent steps. We assumed that $\varphi(x)$ is
a general fourth-order polynomial and we ran a global search finding
its coefficient and the visibility parameter ${\cal V}$. Afterwards
we performed the local optimization with unconstrained values of the
discretized $\varphi(x)$, starting from the result obtained using
the global search. 

\textbf{Results uncertainties.}
To account for the uncertainty of the empirical HSP in the phase retrieval
procedure, we applied the Monte-Carlo approach. We repeated the phase
retrieval procedure 5000 times, each time randomizing $|\Psi(x,x')|^{2}$,
$|\psi_{u}(x)|$ and $|\psi_{r}(x)|$ by drawing the initial counts
values at each pixel from the corresponding Poissonian distributions.
In each realization we obtained $|\Psi_{\textrm{rec}}^{(\varphi)}(x,x')|^{2}$
and the corresponding vector of phase. The Monte-Carlo approach resulted
in mean reconstructed HSP presented in Fig.~4b and the phase profiles
whose mean and standard deviation, after unifing their convexities
and constant phase offset, are presented in Fig.~4c. 
}
\bibliographystyle{apalike}

\footnotesize{
\subsection*{Acknowledgements}

We acknowledge insightful comments and discussion about the work and the
manuscript with M. Barbieri, R. \L apkiewicz, M. J. Padgett and A.
Zeilinger. This project was financed by the National Science Centre
No. DEC-2013/09/N/ST2/02229 and DEC-2011/03/D/ST2/01941. R.C. was supported by Foundation for
Polish Science. M.J and K.B. were supported by the European Commission
under the FP7 IP project SIQS (Grant agreement no. 600645) co-financed
by the Polish Ministry of Science and Higher Education. 

\subsection*{Author Contributions}

W.W. triggered the research and proposed the idea of wavefunction phase retrieval. R.C. designed and programmed the experiment, developed HSP methods, analyzed the data and prepared figures. M.J. built a setup and performed the measurements. R.C. and M.J. wrote the manuscript assisted by W.W. and K.B who supervised the work.
}

\end{document}